# Beyond Demographics: BIM Engagement and Job Satisfaction Among AEC Professionals — A Machine Learning Pilot Study

***Sharareh Mirzaei***
***Gerald May Dept. of Civil, Construction and Environmental Engineering, University of New Mexico;***
***shmirzaei@unm.edu***

***SUMMARY***: *Building Information Modeling (BIM) has transformed workflows across the Architecture, Engineering, and Construction (AEC) industry, yet its relationship with employee job satisfaction remains insufficiently understood. This pilot study investigates whether BIM engagement or demographic characteristics better predict job satisfaction among AEC professionals. Survey responses from 104 participants were analyzed using Spearman rank correlations, logistic regression, and Classification and Regression Tree (CART) modeling. A 27-item Job Satisfaction Index demonstrated excellent internal reliability (Cronbach's α = 0.937). Across all analytical approaches, BIM engagement emerged as a stronger predictor of job satisfaction than demographic factors. Specifically, the proportion of project work completed using BIM was the only significant predictor of job satisfaction, whereas age, gender, education level, and professional experience showed no significant relationships. The CART analysis further identified BIM project involvement as the primary factor associated with higher job satisfaction. These findings suggest that the extent of BIM integration in professional practice may play a more important role in shaping employee satisfaction than individual demographic characteristics. The study contributes to the growing literature on human–technology interactions in the AEC sector and provides preliminary evidence to support strategies that promote deeper BIM adoption.*

# 1. INTRODUCTION

The increasing digitalization of the Architecture, Engineering, and Construction (AEC) industry has made human–technology interaction a central component of modern professional practice. As Building Information Modeling (BIM) becomes embedded in daily workflows, it shapes not only project coordination, information exchange, visualization, and decision-making, but also how professionals experience their work. BIM has been widely recognized for improving project coordination, information management, and project-level outcomes (Azhar, 2011; Bryde et al., 2013; Eastman et al., 2011). However, the benefits of BIM are not produced by software alone. They depend on human behavior, perceived usefulness, ease of use, organizational support, workflow fit, training, and professional expectations (Davis, 1989; Venkatesh et al., 2003). Therefore, understanding how BIM engagement relates to job satisfaction is essential for developing digital work environments that are both technologically effective and human centered.

The human side of BIM-enabled work remains underexplored compared with the technical and project-performance dimensions of BIM. Much of the BIM literature emphasizes system performance, clash detection, cost and schedule control, productivity, interoperability, and coordination benefits, while fewer studies examine how workers experience BIM-driven workflows after adoption (Azhar, 2011; Bryde et al., 2013; Succar, 2009). Existing research shows that BIM adoption is influenced by technological, organizational, managerial, and human factors such as training, readiness, resistance to change, and perceived value (Hyarat et al., 2022; Van Tam et al., 2021). Yet less is known about whether satisfaction is shaped more by who professionals are, such as age, education, experience, or role, or by how deeply BIM is integrated into their daily project work.

Prior studies on BIM user satisfaction suggest that organizational support, BIM performance, managerial involvement, and project complexity may influence how professionals experience BIM-enabled work. Organizational and managerial support can strengthen BIM user satisfaction by providing technical resources, emotional support, training, and clearer implementation guidance (Hua et al., 2024; Jiang et al., 2021). BIM performance dimensions, including accuracy, information integration, and functional usefulness, may also affect satisfaction because they shape whether professionals experience BIM as helpful, efficient, and aligned with project needs (Jiang et al., 2021). In addition, project complexity may moderate these relationships, with BIM performance becoming more important in complex projects and managerial support becoming more visible in less complex project settings (Jiang et al., 2021). Comparative research further suggests that BIM users may experience different workplace outcomes than non-BIM users, including differences in career advancement, work-life balance, skill development, and workplace experience (Inguva, 2014). These findings indicate that BIM engagement should be understood not only as software exposure, but as part of the broader socio-technical work environment (Peng et al., 2025).

This distinction is increasingly important as BIM expands into AI-assisted modeling, automated layout tools, digital twins, and immersive platforms such as augmented and virtual reality. These systems create new forms of interaction among professionals, models, project data, and automated decision-support tools (Noghabaei et al., 2020; Yigitbas et al., 2023; Su et al., 2025). If digital transformation is to improve both project performance and workplace well-being, organizations need evidence about the human factors associated with positive work experiences. This pilot study addresses that need by examining whether BIM engagement or demographic characteristics show stronger associations with job satisfaction among AEC professionals.

Accordingly, this study investigates job satisfaction among AEC professionals by comparing demographic characteristics with BIM engagement variables. Demographic characteristics such as age, education level, years of professional experience, and BIM experience are commonly considered in technology adoption research because they may influence familiarity, confidence, and readiness to use digital tools. However, in BIM-enabled work environments, satisfaction may depend more on the depth of BIM integration in project activities than on demographic profile alone. This study, therefore, examines whether the percentage of project work completed using BIM, frequency of BIM use for collaboration, and years of BIM experience are associated with job satisfaction, and whether these BIM engagement variables provide stronger explanatory value than demographic characteristics.

# 2. LITERATURE REVIEW

## 2.1. Human–Technology Interaction in AEC

Human–technology interaction in the AEC industry has evolved rapidly as technology becomes increasingly integrated with immersive visualization and extended-reality environments. Current research shows that BIM is shifting from a static coordination tool to a multisensory, interactive platform that supports VR, AR, MR, scan-to-BIM pipelines, and gesture-based control, enabling richer communication and spatial understanding across the building lifecycle (Zhang et al., 2025; Banfi & Previtali 2021; Alizadehsalehi et al., 2020; Yigitbas et al., 2023). These interaction modalities enhance model accessibility, contextualize design intent, and connect digital information directly to field activities. At the same time, BIM-XR ecosystems introduce new cognitive demands on users, making human-centered design essential for ensuring that digital environments support daily work. Research in other technology-mediated domains similarly emphasizes that, as automation expands, tools must be grounded in users' actual information needs and must balance automated assistance with meaningful human control to avoid overwhelming users (Nezhad et al., 2024). Studies emphasize that user experience, ergonomics, and interface design strongly influence how effectively professionals interact with BIM-driven systems, reinforcing the need to approach BIM as a socio-technical environment rather than a purely software-based workflow (Peng et al., 2025).

Despite notable advancements, meaningful human-technology-interaction in BIM-enabled environments is still constrained by usability limitations, interoperability issues, and organizational factors that shape worker experience (Peng et al., 2025, Banfi and Previtali, 2021, Yigitbas et al., 2023). Recent research on e-leadership in technology-mediated environments further shows that digital communication, trust formation, and leader–member interaction play central roles in shaping employee engagement and performance within virtual teams (Eslamdoust et al., 2024). Research consistently identifies that without addressing user-centered concerns, such as adequate training, workflow alignment, and reduction of data friction technological gains fail to translate into real practice improvements. Emerging modalities such as gesture interfaces(Park et al., 2022), BIM-XR collaboration(Zhang et al., 2025; Alizadehsalehi et al., 2020), scan-to-BIM digitization(Yigitbas et al., 2023), gamified learning(Feng et al., 2021), and AI-driven visualization(Fernandes et al., 2024) demonstrate strong potential to improve safety, communication, and team coordination, yet most evidence comes from isolated prototypes rather than long-term field implementations(Alizadehsalehi et al., 2020; Peng et al., 2025). These gaps highlight the growing need to examine how different people across ages, roles, experience levels, and digital competencies engage with BIM, and how these interactions shape acceptance, satisfaction, and workplace outcomes. Understanding these human differences is essential for developing future technologies, policies, and training programs that support equitable and effective digital transformation in AEC.

## 2.2. Demographic Factors in Human–Technology Interaction

Demographic characteristics significantly shape how individuals adopt, perceive, and interact with digital technologies. Foundational technology acceptance research shows that perceived usefulness and perceived ease of use influence users' attitudes and intentions toward technology adoption (Davis, 1989). Extending this perspective, the Unified Theory of Acceptance and Use of Technology identifies age, gender, experience, and voluntariness of use as important moderators of technology acceptance and usage behavior (Venkatesh et al., 2003). Prior studies have also shown that age and gender can influence technology adoption decisions and the role of social influence in technology use (Morris & Venkatesh, 2000; Venkatesh & Morris, 2000). In addition, education and digital skill differences shape how effectively individuals use and benefit from digital systems, indicating that technology interaction is influenced not only by access but also by users' knowledge, confidence, and skill level (Hargittai, 2002). Understanding these variations is critical for designing inclusive and effective technological environments, especially as AEC workplaces continue to digitalize and rely more heavily on BIM-enabled coordination, communication, and information-sharing systems (Mirzaei et al., 2026).

Age remains a major determinant of technology interaction. Older adults often encounter barriers, including cost, limited technology fluency, and heightened privacy concerns, which collectively slow adoption compared to younger generations (Wilson, 2018). In contrast, younger individuals typically demonstrate more positive attitudes toward digital tools and artificial intelligence due to greater familiarity and comfort with digital environments (Grassini et al., 2025). Education similarly plays a substantial role; higher educational attainment is consistently associated with increased technology adoption and more sophisticated usage patterns (Kiburu et al., 2023). Within digital service contexts, education has been shown to moderate the relationship between perceived usefulness and behavioral intention, emphasizing its importance in technology acceptance models

(Abu-Shanab, 2021).

Gender and socioeconomic factors also contribute to differences in technology interaction. Studies report that men often exhibit higher technology acceptance and interest (Ferizaj et al., 2023), although gender does not universally predict behavioral intentions across all domains; for instance, it showed no significant effect on e-government adoption (Abu-Shanab, 2021). Income, occupation, and geographic location further influence technology engagement, with higher income and certain occupations providing greater access to digital tools and resources, particularly among older adults (Wilson, 2018). Beyond demographics, psychological attributes such as openness and agreeableness have been shown to shape attitudes toward AI and technology use(Grassini et al., 2025), underscoring that technology adoption is multifaceted and shaped by both personal and contextual factors.

## 2.3. Digital Engagement, and its Influence on Acceptance and Satisfaction

User satisfaction with digital technologies is strongly shaped by familiarity, engagement quality, and the experience of digital touchpoints. Studies show that users who are more familiar with digital tools and applications tend to report higher satisfaction, as familiarity enhances confidence, reduces cognitive effort, and supports smoother interaction (Kumari, 2024). Digital engagement also acts as a key mediator between user interaction and satisfaction; for example, in the Alodokter health application, engagement significantly improved satisfaction, highlighting the importance of interactive design and high-quality informational content (Ugut et al., 2025). Experiences with digital touchpoints such as applications or service interfaces further determine satisfaction levels and contribute to long-term loyalty (Nanta et al., 2025). However, engagement alone does not guarantee positive user experiences; the quality and relevance of content, as well as alignment with user expectations, are essential for sustaining satisfaction over time (Jo and Ahn, 2024). This underscores the need for digital services that balance engagement with meaningful, user-centered design.

Satisfaction with digital tools plays a critical role in shaping job satisfaction by enhancing productivity, autonomy, and overall well-being. Effective digital systems improve task coordination and collaboration, leading employees to experience greater engagement and fulfillment in their work (Kazmi and Irshad, 2025). These tools can also increase autonomy and flexibility, support core psychological needs, and contribute to more positive workplace experiences (Mukherjee and Gopal, 2024). However, poorly integrated or overly demanding digital systems may introduce time pressure, stress, or job insecurity, which can undermine satisfaction and highlight the need for thoughtful design and organizational support (Bolli and Pusterla 2022). Within BIM-enabled AEC environments, digital engagement is not limited to how often professionals use a tool. It also reflects how deeply the tool is embedded in project workflows, information exchange, collaboration practices, and decision-making processes. Prior to BIM-focused research suggests that user satisfaction is shaped by the interaction between technological performance, organizational support, and project conditions. Organizational support, including emotional and instrumental support, can improve BIM user satisfaction by helping professionals access resources, develop skills, and apply BIM more effectively in daily work (Hua et al., 2024). Similarly, managerial support can influence satisfaction by clarifying expectations, reducing implementation uncertainty, and helping users connect BIM processes to project goals (Jiang et al., 2021).

BIM performance dimensions are also important because they influence whether professionals perceive BIM as beneficial or burdensome. When BIM supports model accuracy, information integration, coordination, and functional usefulness, it may improve work experience by reducing information fragmentation and supporting clearer communication across project teams (Jiang et al., 2021). However, when BIM is poorly integrated or unsupported by organizational processes, it may increase coordination effort, learning demands, frustration, and role ambiguity. Therefore, BIM engagement should be understood as a socio-technical condition that depends on both the technology itself and the organizational environment in which the technology is used. Recent advances in BIM-enabled digital-twin frameworks have demonstrated substantial reductions in estimating effort and improved probabilistic schedule control under real construction conditions, suggesting that deeply integrated BIM environments can meaningfully reshape how professionals experience their daily work (Khoshkonesh et al., 2026)

Project complexity may further influence the relationship between BIM engagement and satisfaction. In complex projects, BIM may provide greater value by improving coordination, supporting information integration, and helping professionals manage uncertainty.

Comparative research also suggests that BIM users and non-BIM users may experience different workplace outcomes. BIM users may report more positive perceptions of career advancement, work-life balance, workplace experience, and skill development than non-BIM users (Inguva, 2014). Although these findings require further validation in contemporary BIM environments, they support the argument that BIM engagement can shape professional experience beyond technical productivity. BIM may influence how workers perceive their roles, growth opportunities, collaboration quality, and contribution to project outcomes.

Overall, the literature indicates that BIM engagement may operate as a human-centered work condition rather than a simple measure of software use. Existing studies have examined BIM user satisfaction, organizational support, BIM performance, and project complexity, but limited empirical work has directly tested whether BIM engagement variables explain job satisfaction more consistently than demographic characteristics. This gap provides the basis for the present pilot study, which compares demographic variables with BIM engagement indicators to examine whether the depth of BIM integration in project work is more closely associated with job satisfaction among AEC professionals.

# 3. METHODOLOGY

## 3.1. Research Design

This study employed a quantitative cross-sectional survey design to investigate the relationship between BIM engagement and job satisfaction among AEC professionals. Cross-sectional survey designs are commonly used to examine relationships among variables at a single point in time and are particularly useful for identifying patterns, associations, and preliminary predictors within a target population (Creswell & Creswell, 2018; Fowler, 2014). Given the limited empirical research examining BIM engagement as a predictor of employee satisfaction, the study was designed as a pilot investigation intended to generate preliminary evidence, assess the feasibility of the research approach, and inform future large-scale studies (Leon et al., 2011; Thabane et al., 2010).

To enhance the robustness of the findings, a multi-method analytical framework was adopted. Three complementary techniques—Spearman rank correlation, binary logistic regression, and Classification and Regression Tree (CART) analysis—were applied to examine relationships from both inferential and predictive perspectives. Spearman rank correlation was used to assess monotonic associations without assuming normality, making it appropriate for ordinal or non-normally distributed survey data (Spearman, 1904). Binary logistic regression was used to evaluate the likelihood of job satisfaction outcomes based on BIM engagement and demographic predictors, following its established use for modeling dichotomous dependent variables (Hosmer et al., 2013). CART analysis was further employed to identify hierarchical decision rules and predictor importance through a non-parametric, tree-based modeling approach (Breiman et al., 1984). The use of multiple analytical approaches enabled methodological triangulation, allowing the study to compare the consistency of findings across different techniques and reduce reliance on the assumptions and limitations associated with any single statistical method (Denzin, 1978; Jick, 1979).

## 3.2. Data Collection and Sample

Data were collected through an online questionnaire administered via Qualtrics and distributed to AEC professionals through professional networks, industry contacts, and academic channels. Participation was voluntary and anonymous.

A total of 119 responses were received. Fifteen responses were excluded because no job satisfaction items were completed, leaving insufficient information for meaningful analysis or imputation. The remaining dataset consisted of 104 respondents, including 88 fully completed surveys and 16 partially completed surveys that were retained following data imputation procedures described in Section 3.4. Consequently, the final analytical sample comprised 104 respondents.

## 3.3. Survey Instrument

The survey instrument consisted of two sections: (1) demographic and BIM engagement characteristics, and (2) job satisfaction measures.

### 3.3.1. Demographic and BIM Engagement Variables

Demographic variables included gender, age group, highest educational attainment, primary professional role, and years of industry experience. BIM engagement was assessed using three indicators:

Years of BIM experience, Frequency of BIM use for collaboration, measured on a five-point ordinal scale ranging from Never to Daily, and Percentage of project work completed using BIM tools, measured on a four-category ordinal scale ranging from Under 15% to Above 60%.

### 3.3.2. Job Satisfaction Measure

Job satisfaction was measured using a 27-item Likert-type scale with responses ranging from 1 (Strongly Disagree) to 5 (Strongly Agree). The instrument captured five dimensions of work experience:

General job meaningfulness, BIM-related skill utilization and task variety, BIM-enabled collaboration, Autonomy and performance feedback, and Technology-related productivity and challenges.

Three negatively worded items were reverse coded to ensure that higher scores consistently reflected more positive job experiences. These items addressed perceived micromanagement through BIM-generated data, role ambiguity associated with BIM processes, and stress related to BIM learning requirements.

A composite Job Satisfaction Index (JSI) was calculated as the arithmetic mean of all 27 items, producing a continuous score ranging from 1 to 5. Internal consistency reliability was assessed using Cronbach's alpha, yielding $\alpha=0.937$, which indicates excellent reliability and supports the use of a single composite measure.

## 3.4. Data Preparation

### 3.4.1. Treatment of Missing Data

Responses containing no completed job satisfaction items (n=15) were removed because reliable imputation was not possible. For the remaining partially completed surveys (n=16), missing values within the job satisfaction scale were addressed using person-mean imputation. Under this approach, each missing item was replaced by the respondent's average score across completed job satisfaction items.

Person-mean imputation was selected because it preserves individual response patterns and has been widely recommended for psychometric scales when missingness is limited. An imputation flag variable was retained to facilitate sensitivity analyses and transparency regarding data treatment.

### 3.4.2. Variable Coding

Categorical variables were numerically encoded prior to analysis. Ordinal variables—including age group, educational attainment, years of professional experience, BIM experience, BIM collaboration frequency, and BIM project involvement—were coded according to their natural ordering.

### 3.4.3. Outcome Variable

The Job Satisfaction Index was used as a continuous outcome measure for correlation analyses. For classification-based modeling, the index was dichotomized using the sample median (3.815) to create two categories:

High job satisfaction (JSI $\geq$3.815; n=55)

Low job satisfaction (JSI <3.815; n=49)

Although dichotomization may reduce statistical information, it was necessary for logistic regression and CART classification analyses. Retaining both continuous and binary representations enabled the use of methods appropriate to each analytical objective.

## 3.5. Statistical Analysis

### 3.5.1. Spearman Rank Correlation Analysis

Spearman's rank-order correlation coefficient ($\rho$) was used to assess bivariate associations between predictor variables and the continuous Job Satisfaction Index. This nonparametric approach was selected because several predictors were measured on ordinal scales and because it does not require assumptions of normality. Correlations were interpreted based on both statistical significance and effect size.

### 3.5.2. Binary Logistic Regression

Binary logistic regression was conducted to evaluate the independent contribution of each predictor while controlling all other variables in the model. Prior to estimation, predictors were standardized using z-score normalization to facilitate comparison of coefficient magnitudes.

Given the relatively small sample size and the number of predictors included, L2 (Ridge) regularization was applied to reduce model overfitting and improve coefficient stability. Odds ratios and corresponding 95% confidence intervals were estimated using bootstrap resampling with 1,000 iterations, providing more reliable interval estimates than asymptotic standard errors under small-sample conditions.

### 3.5.3. Classification and Regression Tree (CART) Analysis

A CART model was developed to identify interpretable decision rules associated with high versus low job satisfaction. Decision trees were selected because they accommodate nonlinear relationships, interactions among predictors, and mixed data types while remaining readily interpretable for applied decision-making.

The dataset was partitioned into training (70%) and testing (30%) subsets using stratified random sampling to preserve class balance. Hyperparameters were optimized through grid search using 10-fold stratified cross-validation, with model selection based on the area under the receiver operating characteristic curve (AUC). Tree depth and minimum leaf size were constrained to reduce overfitting and improve model generalizability.

Model performance was evaluated using accuracy, sensitivity, specificity, F1-score, and AUC. To assess stability, the final model was further evaluated using 10-fold cross-validation on the full dataset. Predictor importance was quantified using reductions in Gini impurity.

## 3.6. Software and Reproducibility

All analyses were performed using Python 3. Data preparation and management were conducted using pandas and NumPy. Statistical analyses were implemented using SciPy and scikit-learn, while visualizations were produced using Matplotlib.

To support transparency and reproducibility, all analytical scripts, model specifications, and the cleaned dataset containing imputation indicators have been archived and are available from the authors upon reasonable request.

# 4. RESULTS

## 4.1. Sample Characteristics

The final analytical sample consisted of 104 AEC professionals. Respondents were predominantly male (67.3%), highly educated, and represented a range of professional roles and experience levels. BIM Specialists constituted the largest professional subgroup, followed by Engineers and Architects. The sample demonstrated substantial BIM engagement, with 74.0% reporting BIM use on more than 60% of their projects and 73.1% using BIM for collaboration at least weekly. Detailed demographic and professional characteristics are presented in Table 1.

*Table 1. Demographic and Professional Characteristics of Respondents (n = 104)*

| Variable | Category | n | % |
|---|---|---|---|
| Gender | | | |
| | Male | 70 | 67.3% |
| | Female | 30 | 28.8% |
| | Prefer not to say | 2 | 1.9% |
| | Non-binary / third gender | 1 | 1.0% |
| Age Group | | | |
| | Under 25 | 9 | 8.7% |
| | 25–34 | 45 | 43.3% |
| | 35–44 | 30 | 28.8% |
| | 45–54 | 16 | 15.4% |
| | 55 and above | 4 | 3.8% |
| Education Level | | | |
| | Secondary education | 6 | 5.8% |
| | Associate's degree | 9 | 8.7% |
| | Bachelor's degree | 41 | 39.4% |
| | Master's degree | 46 | 44.2% |
| | Doctoral degree | 2 | 1.9% |
| Professional Role (multi-select) | | | |
| | BIM Specialist | 43 | 41.3% |
| | Engineer | 33 | 31.7% |
| | Architect | 31 | 29.8% |
| | Project Manager | 11 | 10.6% |
| | Contractor | 7 | 6.7% |
| | Other | 11 | 10.6% |
| Years of Experience | | | |
| | Less than 1 year | 2 | 1.9% |
| | 1–3 years | 18 | 17.3% |
| | 4–6 years | 25 | 24.0% |
| | 7–10 years | 18 | 17.3% |
| | More than 10 years | 41 | 39.4% |
| BIM Experience | | | |
| | Less than 1 year | 10 | 9.6% |
| | 1–3 years | 19 | 18.3% |
| | 4–6 years | 23 | 22.1% |
| | 7–10 years | 25 | 24.0% |
| | More than 10 years | 27 | 26.0% |
| BIM % of Projects | | | |
| | Under 15% | 9 | 8.7% |
| | 15–30% | 9 | 8.7% |
| | 31–60% | 9 | 8.7% |
| | Above 60% | 77 | 74.0% |
| BIM for Collaboration | | | |
| | Never | 6 | 5.8% |
| | Rarely | 14 | 13.5% |
| | Monthly | 7 | 6.7% |
| | Weekly | 35 | 33.7% |
| | Daily | 41 | 39.4 |

## 4.2. Job Satisfaction Index

### 4.2.1. Reliability and Descriptive Statistics

The 27-item Job Satisfaction Index (JSI) demonstrated excellent internal consistency (Cronbach's $\alpha = 0.937$), supporting its use as a composite measure. The mean JSI score was 3.71 ($SD = 0.52$), indicating generally positive job satisfaction among respondents. Scores ranged from 2.30 to 5.00, with a median of 3.82.

Item-level analysis revealed that respondents reported the highest levels of satisfaction regarding role clarity, workplace relationships, perceived importance of their work, and overall job enjoyment. In contrast, BIM-related stress reduction and deadline management received the lowest ratings, suggesting that while BIM may support productivity and collaboration, its benefits in reducing workplace stress remain limited.

## 4.3. Spearman Correlation Analysis

Spearman rank correlations were computed to examine bivariate associations between predictor variables and the continuous JSI (Table 2). Of the eight predictors examined, only BIM project involvement demonstrated a statistically significant positive association with job satisfaction ($rho = 0.212$, $p = .030$). Professionals who reported using BIM on a larger proportion of their projects tended to report higher levels of job satisfaction.

**Table 2. Spearman Rank Correlations between Predictor Variables and Job Satisfaction Index ($n = 104$)**

| Predictor | *rho* | p-value | Interpretation |
|---|---|---|---|
| **BIM Engagement Variables** | | | |
| BIM % of Projects | **0.212** | **.030*** | Positive, statistically significant — small effect |
| BIM Collaboration | 0.140 | .155 | Positive trend, not significant |
| BIM Experience | −0.023 | .817 | No meaningful association |
| **Demographic & Occupational Variables** | | | |
| Professional Role | 0.196 | .046 | Marginal positive trend; interpret cautiously given exploratory context |
| Gender (Male) | 0.086 | .385 | Not significant |
| Age Group | −0.100 | .313 | Not significant |
| Education | −0.081 | .414 | Not significant |
| Years of Experience | −0.121 | .220 | Not significant |

*Note. rho = Spearman rank-order correlation coefficient. p-values are two-tailed. * $p < .05$. Predictors are grouped by variable types. Professional Role was treated as a single ordinal variable reflecting the primary occupational category. The Kruskal-Walli's test confirmed no significant overall group difference in satisfaction across role categories ($H = 5.47$, $p = .362$); the marginal Spearman trend should therefore be interpreted with caution.*

No significant relationships were observed for demographic variables, including age group (rho = −0.100, p = .313), gender (rho = 0.086, p = .385), educational attainment (rho = −0.081, p = .414), or years of professional experience (rho = −0.121, p = .220). BIM experience (rho = −0.023, p = .817) and BIM collaboration frequency (rho = 0.140, p = .155) were also not significantly associated with job satisfaction. Professional role, treated as a single categorical variable, showed a marginal positive trend (rho = 0.196, p = .046) but did not reach conventional thresholds for strong significance after accounting for the exploratory nature of the analysis.

Overall, the correlation analysis suggests that the depth of BIM integration within project workflows is more closely related to job satisfaction than demographic or occupational characteristics.

**Figure 1. Spearman Rank Correlations with Job Satisfaction Index**

### 4.4. Logistic Regression Analysis

The results of the binary logistic regression model are presented in Table 3. The model achieved an overall classification accuracy of 66.3%.

No predictor achieved statistical significance based on bootstrap-derived 95% confidence intervals, likely reflecting the limited statistical power of the current sample size relative to the number of predictors. Nevertheless, the direction of effects largely aligned with the correlation analysis. BIM project involvement exhibited the strongest positive association with high job satisfaction (OR = 1.41, 95% CI [0.89–3.19]), followed by professional role (OR = 1.34) and gender (OR = 1.38). Conversely, years of professional experience (OR = 0.72) and BIM experience (OR = 0.80) demonstrated negative associations with satisfaction, suggesting that longer tenure alone does not guarantee higher satisfaction in BIM-enabled environments.

The consistency of coefficient directions with the bivariate findings provides additional support for the importance of BIM engagement, even in the absence of formal statistical significance.

**Table 3. Exploratory Logistic Regression: Predictors of High Job Satisfaction (n = 104)**

| Predictor | Coef | OR | 95% Bootstrap CI | Direction |
|---|---|---|---|---|
| **BIM Engagement Variables** | | | | |
| BIM % of Projects | 0.344 | **1.41** | [0.89 – 3.18] | Positive (strongest) |
| BIM Experience | −0.227 | 0.80 | [0.40 – 1.54] | Negative |
| BIM Collaboration | 0.007 | 1.01 | [0.56 – 1.61] | Positive |
| **Demographic Variables** | | | | |

| Years of Experience | −0.331 | 0.72 | [0.32 – 1.31] | Negative |
|---|---|---|---|---|
| Age Group | 0.169 | 1.18 | [0.73 – 2.09] | Positive |
| Gender (Male) | 0.322 | 1.38 | [0.91 – 2.26] | Positive |
| Education | −0.054 | 0.95 | [0.57 – 1.79] | Negative |
| Professional Role | 0.291 | 1.34 | [0.90 – 2.23] | Positive |
| *Model accuracy: 66.3% \| No predictor achieved significance (all bootstrap 95% CIs cross 1.00) \| L2 regularisation applied* | | | | |

*Note. Coef = standardised logistic regression coefficient; OR = odds ratio; CI = confidence interval estimated using 1,000 bootstrap iterations. OR > 1 indicates higher odds of high job satisfaction; OR < 1 indicates lower odds. All confidence intervals cross 1.00, indicating no predictor reached statistical significance. Results are directional and hypothesis-generating only.*

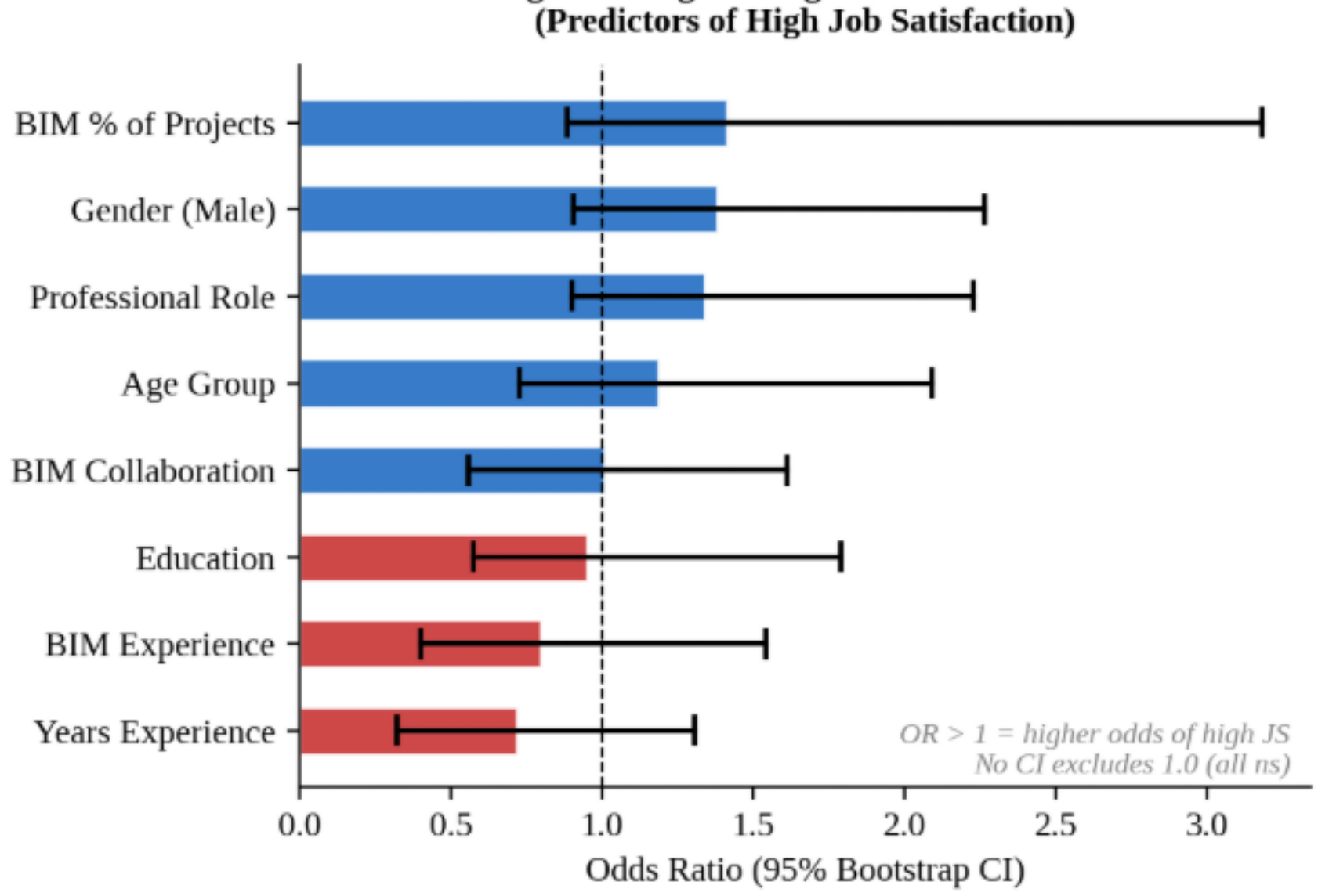

## 4.5. Classification and Regression Tree (CART) Analysis

### 4.5.1. Model Performance

The optimized CART model (max_depth = 2, min_samples_leaf = 10) achieved a test-set accuracy of 50.0% and an AUC of 0.529. Ten-fold cross-validation yielded a mean AUC of 0.592 (SD = 0.143), indicating modest predictive capability above random classification. Although predictive performance was limited, the model remained useful for identifying influential variables and interpretable decision rules consistent with the other analytical approaches

### 4.5.2. Variable Importance and Decision Rules

Variable importance analysis identified BIM project involvement as the dominant predictor of job satisfaction (Gini importance = 0.500), followed by professional role (0.500). All remaining variables contributed negligible importance to the final model.

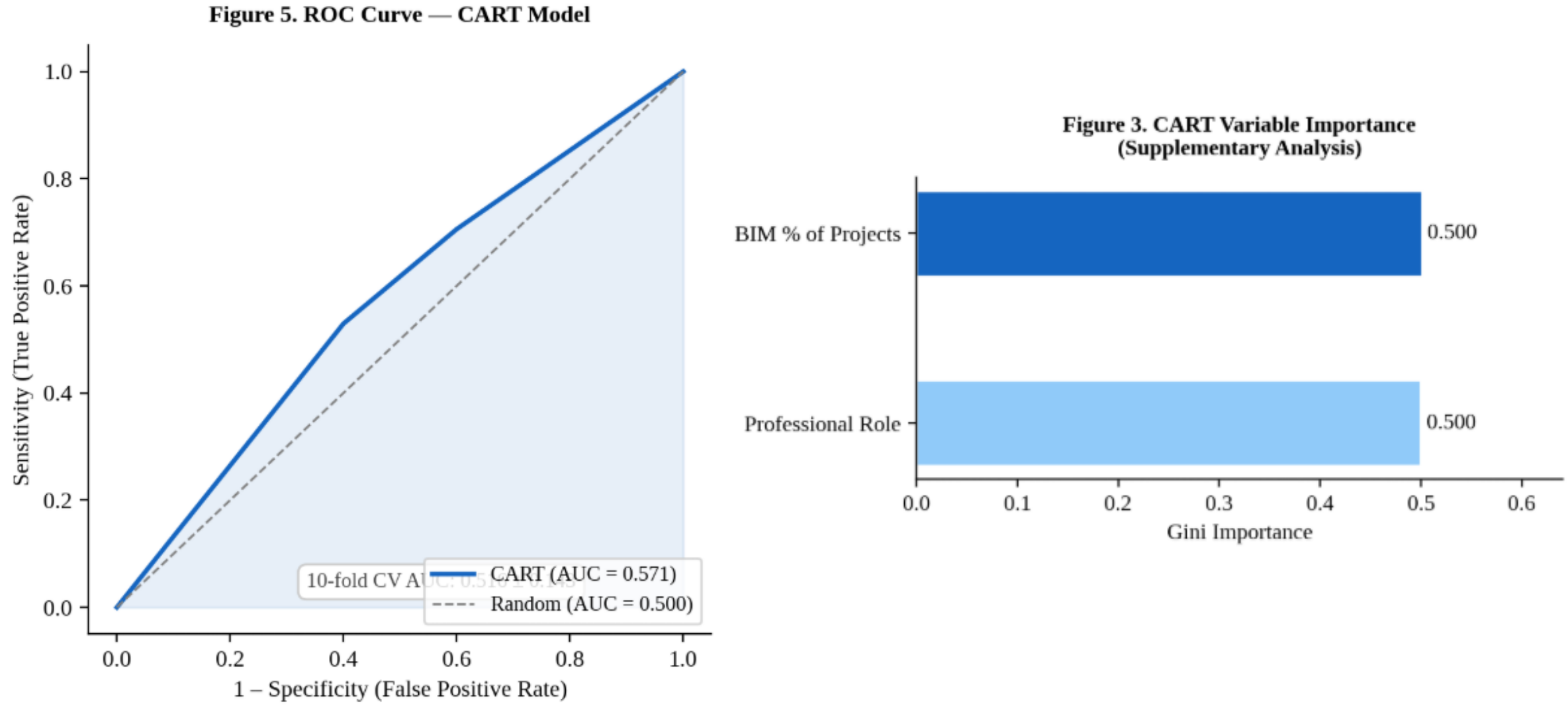

**Figure 5. ROC Curve — CART Model**

**Figure 3. CART Variable Importance (Supplementary Analysis)**

Importantly, professional role was entered as a single categorical variable reflecting each respondent's primary occupational category, rather than as separate binary indicators. This treatment is consistent with its non-significant overall effect in the Kruskal-Wallis test ($H = 5.47$, $p = .362$), which confirmed that no individual role category produced meaningfully different satisfaction levels when compared as a group. Its appearance in the CART model therefore reflects a secondary interaction with BIM project involvement rather than an independent effect on satisfaction.

**Figure 4. Pruned Decision Tree for Job Satisfaction Classification (max_depth = 2, min_samples_leaf = 10, n = 104)**

The primary split in the decision tree was based on BIM project involvement. Respondents using BIM on more than 60% of their projects were generally classified into the high-satisfaction group, whereas those with lower levels of BIM integration were classified into the low-satisfaction group. A secondary split based on professional role further refined predictions among respondents with lower BIM project involvement, though this interaction should be interpreted cautiously given the sample size.

These findings reinforce the results of the correlation and regression analyses by highlighting BIM project involvement as the principal factor differentiating job satisfaction levels.

## 4.6. Cross-Method Synthesis

Comparison of findings across Spearman correlation, logistic regression, and CART analyses revealed a consistent pattern. BIM project involvement emerged as the strongest predictor of job satisfaction across all three analytical approaches (Spearman rho = 0.212, p = .030; OR = 1.41; Gini importance = 0.500). In contrast, demographic characteristics — including age, gender, education, and years of experience — showed limited and non-significant explanatory value across all methods.

Professional role was examined as a unified categorical variable. Although a marginal trend emerged in the Spearman analysis (rho = 0.196, p = .046), the Kruskal-Wallis test confirmed no significant overall difference in satisfaction across role groups (H = 5.47, p = .362). This distinction is methodologically important: treating role as separate binary indicators, as done in some prior machine learning studies, risks producing spurious role-specific effects that do not reflect genuine group differences.

The convergence of results across multiple analytical methods, combined with the correction of role treatment, strengthens confidence in the central finding of this study: job satisfaction among AEC professionals is more closely associated with the extent of BIM integration in daily project work than with demographic or occupational characteristics. Although the observed effects were modest in magnitude — consistent with expectations for an exploratory pilot study — their consistency across methods suggests that BIM engagement represents a meaningful and theoretically grounded avenue for future investigation using larger, longitudinal samples.

# 5. DISCUSSION

This study investigated whether BIM engagement or demographic characteristics better explain job satisfaction among AEC professionals. Across three complementary analytical approaches — Spearman correlation, logistic regression, and CART analysis — a consistent pattern emerged: BIM project involvement was more strongly associated with job satisfaction than demographic or role-related characteristics. Although the observed effects were modest, their convergence across methods suggests that the extent to which BIM is embedded within daily project work plays a meaningful role in shaping employee experiences. Critically, professional role was examined as a unified categorical variable in this study, and no significant overall difference in satisfaction was found across role groups (Kruskal–Wallis H = 5.47, p = .362), underscoring that occupational title alone does not determine how BIM shapes workplace experiences.

## 5.1. BIM Engagement and Job Satisfaction

The most robust finding of this study was the positive association between BIM project involvement and job satisfaction (rho = 0.212, p = .030; OR = 1.41). Among all predictors examined, the proportion of project work completed using BIM was the only variable that demonstrated a statistically significant bivariate relationship with satisfaction and emerged as the most influential predictor across all three analytical methods.

This finding aligns with recent BIM job satisfaction research suggesting that BIM-enabled work environments influence employee satisfaction not simply through technology use, but through the quality of work experiences created by BIM-mediated collaboration, feedback, meaningful engagement, and reduction of workflow pain points (Mirzaei et al., 2026). BIM functions not only as a technical modeling platform but also as an organizational infrastructure that facilitates information exchange, interdisciplinary collaboration, and workflow transparency. Professionals who work within highly BIM-integrated environments may therefore experience greater efficiency, reduced ambiguity, and stronger perceptions of task accomplishment — all of which are associated with positive workplace outcomes (Mirzaei et al., 2026).

The CART analysis additionally suggested that the relationship between BIM project involvement and satisfaction may interact with professional role. Among respondents using BIM on more than 60% of their projects, satisfaction was generally higher, regardless of occupational background. This reinforces the argument that it is the depth of BIM integration in daily work — rather than the specific role a professional occupies — that most directly shapes workplace experiences.

## 5.2. BIM Experience and Satisfaction Over Time

A counterintuitive finding emerged in both the logistic regression and CART analyses: BIM experience showed a slightly negative association with job satisfaction (OR = 0.80), suggesting that longer BIM tenure does not necessarily translate into higher satisfaction. One possible explanation is that advanced users become increasingly aware of software limitations, interoperability challenges, and organizational inefficiencies that are less visible to newer users. Highly experienced BIM professionals may also hold higher expectations for workflow integration and collaboration quality, making them more sensitive to gaps between BIM's theoretical potential and its practical implementation within their organizations.

Earlier computer user satisfaction research has similarly suggested that regular users may report lower satisfaction than irregular users, possibly because frequent interaction exposes them to more system shortcomings and unmet needs (Yaverbaum, 1988). This interpretation is also consistent with expectation-confirmation theory, which suggests that post-adoption satisfaction depends on whether actual system performance meets users' expectations (Bhattacherjee, 2001).

## 5.3. Collaboration and BIM-Centric Workflows

Although BIM collaboration frequency did not emerge as a statistically significant predictor in this study (rho = 0.140, p = .155), its consistently positive direction across all three analyses suggests that collaborative BIM workflows may contribute indirectly to job satisfaction. Prior BIM research has emphasized that the value of BIM extends beyond technical modeling capabilities and is strongly linked to its capacity to support communication, coordination, and information sharing among project stakeholders (Mirzaei et al., 2026).

The consistency of this directional trend, even without formal significance, is noteworthy in an exploratory context. It suggests that BIM's contribution to satisfaction may be partly mediated by the quality of collaboration it enables, rather than by BIM use alone. This interpretation is supported by Mirzaei et al. (2026), who found that collaboration was one of the strongest contributors to BIM-related job satisfaction and argued that BIM's collaborative capacity is a central mechanism through which BIM enhances work experience.

## 5.4. Professional Role and Job Satisfaction

The professional role was treated as a single categorical variable in this study, consistent with its theoretical status as a unified occupational characteristic. The Kruskal–Wallis test confirmed no statistically significant overall difference in job satisfaction across role groups (H = 5.47, p = .362), indicating that BIM's positive associations with satisfaction are not concentrated in any particular occupational category.

A marginal positive trend was observed for professional role in the Spearman analysis (rho = 0.196, p = .046), with BIM Specialists exhibiting the highest mean satisfaction (M = 3.82) compared to other groups. However, this trend should be interpreted with caution. BIM Specialists typically operate within highly digitalized project environments and possess greater technical proficiency, making it difficult to isolate role identity from BIM engagement intensity. The observed trend therefore likely reflects the confounding influence of deeper BIM integration rather than an independent effect of occupational title. This interpretation is consistent with the broader finding that BIM engagement depth — not who you are — is the more important explanatory factor.

## 5.5. Limited Influence of Demographic Characteristics

A notable finding was the absence of meaningful relationships between job satisfaction and demographic variables including age (rho = −0.100, p = .313), gender (rho = 0.086, p = .385), education (rho = −0.081, p = .414), and years of professional experience (rho = −0.121, p = .220). These results suggest that satisfaction differences are less attributable to individual characteristics than to the nature of the work environment and the degree of technological integration.

This finding contrasts with portions of the technology adoption literature that report demographic differences in attitudes toward digital tools. However, it is consistent with the argument that once technology becomes embedded within routine work processes, organizational and workflow factors exert a stronger influence on employee outcomes than personal characteristics. Mirzaei et al. (2026) similarly argued that BIM-related job satisfaction depends less on technology use alone and more on the work conditions BIM reshapes, including collaboration, meaningful engagement, constructive feedback, and pain-point reduction. From a managerial perspective, this suggests that efforts to improve employee experiences through BIM implementation may yield broadly comparable benefits across diverse workforce groups, regardless of age, gender, or educational background.

## 5.6. Implications for Practice

The findings carry out several practical implications for AEC organizations. First, the results suggest that deeper BIM integration may yield benefits extending beyond project performance and operational efficiency to include employee well-being and workplace satisfaction. Organizations should therefore view BIM implementation not solely as a technical initiative but also as a component of workforce development and organizational improvement.

Second, the findings indicate that meaningful integration is likely more important than nominal adoption. Simply possessing BIM capabilities may be insufficient; the benefits appear strongest when BIM is embedded across a substantial proportion of project activities and used as a central platform for coordination and communication. Investments in training, workflow integration, and cross-disciplinary collaboration may therefore yield greater workforce satisfaction returns than technology acquisition alone.

Third, because no significant demographic differences in BIM-related satisfaction were observed, organizations need not tailor BIM integration strategies by age group, gender, or educational background. A uniformly high-quality BIM environment appears likely to benefit all workforce segments comparably.

## 5.7. Methodological Contributions

This study contributes methodologically in two ways. First, it demonstrates the value of combining correlational, regression-based, and machine-learning approaches within a single exploratory framework. Despite their differing assumptions and analytical objectives, all three methods converged on BIM project involvement as the dominant predictor of job satisfaction, strengthening confidence in the observed relationships and providing a foundation for future confirmatory research.

Second, this study illustrates an important analytical caution for machine learning studies involving categorical occupational variables: treating multi-category role variables as separate binary indicators can produce spurious predictor-specific findings that do not reflect genuine group-level differences. More broadly, the integration of AI and machine learning techniques into domain-specific research workflows requires careful model selection and rigorous consideration of data quality and practical constraints (Omidmand et al., 2025). By treating professional role as a unified categorical variable and validating its overall effect with a non-parametric group test, this study demonstrates a more methodologically defensible approach.

As BIM research continues to expand, greater attention to employee experiences and organizational outcomes — alongside the traditional emphasis on productivity and project performance — will be essential for developing a complete understanding of how digital transformation shapes the construction workforce.

# 6. Limitations and Future Research

Several limitations should be considered when interpreting the findings of this study. First, the analytic sample of 104 respondents, while sufficient for exploratory purposes, limits statistical power and constrains the number of predictors that can be examined simultaneously, as reflected in the wide confidence intervals observed in the logistic regression. The cross-sectional design further prevents causal inference; it remains unclear whether deeper BIM engagement leads to higher satisfaction or whether more satisfied professionals are simply more inclined to adopt BIM extensively. All variables were assessed through self-report, introducing potential biases related to social desirability and common method variance, and the recovery of 16 partially completed responses through person-mean imputation may have introduced a mild upward bias, as imputed respondents exhibited slightly higher mean satisfaction than complete respondents ($M = 3.91$ vs. $M = 3.68$). The sample was also predominantly male (67.3%), highly educated, and heavily skewed toward high BIM engagement (74% using BIM on more than 60% of projects), limiting generalizability to broader AEC populations with lower digital maturity. Finally, the CART model achieved modest predictive performance (test-set AUC = 0.529; 10-fold CV AUC = 0.592), and the sensitivity of tree structures to sample composition at small $n$ reinforces that all findings should be interpreted as exploratory and hypothesis-generating rather than confirmatory.

Future research should prioritize replication with larger, more diverse samples of 300–500 professionals to enable confirmatory structural equation modeling and adequately powered multivariate analyses. Longitudinal designs tracking the same professionals over time — particularly before and after major BIM implementation initiatives — are needed to establish causal direction and examine how satisfaction trajectories evolve as BIM expertise deepens. The counterintuitive negative association between BIM experience and satisfaction observed here (OR = 0.80) warrants specific investigation, as advanced users may face distinct challenges related to interoperability, organizational inefficiencies, and unmet expectations that are less visible to newer practitioners. Future studies should also examine potential mediating mechanisms — such as perceived autonomy, workflow clarity, and collaboration quality — and moderating factors such as organizational BIM maturity and leadership support. Underrepresented groups including women, field-based workers, and professionals in low-BIM environments should be specifically recruited to test whether the patterns observed here generalize beyond BIM-engaged, professionally trained AEC workers. Finally, as BIM increasingly integrates with artificial intelligence, digital twins, and automation tools, future research should extend the human-centered lens of this study to examine how these emerging digital environments shape employee satisfaction and well-being in AEC practice. Such integration also raises organizational adoption challenges — including workforce skill gaps, data governance, and ethical considerations — that recur across AI-enabled experimental and innovation workflows in other sectors (Omidmand & Ataei, 2026) and that AEC organizations will likewise need to manage.

## 7. CONCLUSION

This pilot study examined whether BIM engagement or demographic characteristics better predict job satisfaction among AEC professionals, using survey data from 104 respondents and a multi-method analytical framework comprising Spearman correlation, logistic regression, and CART analysis. Across all three approaches, the proportion of project work completed using BIM emerged as the only statistically significant predictor of job satisfaction (rho = 0.212, $p = .030$; OR = 1.41), while demographic characteristics — including age, gender, educational attainment, and years of professional experience — demonstrated no meaningful explanatory value. Professional role, examined as a unified categorical variable, likewise showed no significant overall effect on satisfaction (Kruskal–Wallis $H = 5.47$, $p = .362$), reinforcing that it is the depth of BIM integration in daily project work, rather than occupational identity or personal background, that most closely shapes how AEC professionals experience their work.

These findings carry both practical and methodological significance. For practitioners and organizations, the results suggest that meaningful BIM implementation — embedding BIM deeply across project workflows and collaborative processes — may yield benefits extending beyond operational efficiency to include employee well-being and workplace satisfaction. Nominal adoption without genuine integration is unlikely to produce these effects. For researchers, this study demonstrates the value of treating multi-category occupational variables as unified constructs rather than separate binary indicators, and illustrates how convergence across correlational, regression-based, and machine-learning methods can strengthen confidence in exploratory findings despite sample size constraints. As a pilot investigation, the study is necessarily limited in its causal scope and generalizability, and future research should employ larger and more diverse samples, longitudinal designs, and mediation analyses to establish the mechanisms through which BIM engagement influences satisfaction. Such work will be essential for understanding how digital transformation can support not only organizational performance but also the workforce well-being of an increasingly data-driven construction industry.

# REFERENCES

Abu-Shanab, E. A. (2021). Demographic factors as determinants of e-government adoption. In Recent developments in individual and organizational adoption of ICTs. IGI Global.

Alizadehsalehi, S., Hadavi, A., & Huang, J. C. (2020). From BIM to extended reality in AEC industry. Automation in Construction, 116, 103254. https://doi.org/10.1016/j.autcon.2020.103254

Azhar, S. (2011). Building information modeling (BIM): Trends, benefits, risks, and challenges for the AEC industry. Leadership and Management in Engineering, 11(3), 241-252. https://doi.org/10.1061/(ASCE)LM.1943-5630.0000127

Banfi, F., & Previtali, M. (2021). Human-computer interaction based on scan-to-BIM models, digital photogrammetry, visual programming language and eXtended Reality (XR). Applied Sciences, 11(13), 6109. https://doi.org/10.3390/app11136109

Bhattacherjee, A. (2001). Understanding information systems continuance: An expectation-confirmation model. MIS Quarterly, 25(3), 351–370. https://doi.org/10.2307/3250921

Bolli, T., & Pusterla, F. (2022). Decomposing the effects of digitalization on workers' job satisfaction. International Review of Economics, 69(2), 263-300. https://doi.org/10.1007/s12232-021-00392-0

Breiman, L., Friedman, J. H., Olshen, R. A., & Stone, C. J. (1984). Classification and regression trees. Wadsworth.

Bryde, D., Broquetas, M., & Volm, J. M. (2013). The project benefits of building information modelling (BIM). International Journal of Project Management, 31(7), 971-980. https://doi.org/10.1016/j.ijproman.2012.12.001

Creswell, J. W., & Creswell, J. D. (2018). Research design: Qualitative, quantitative, and mixed methods approaches (5th ed.). SAGE.

Davis, F. D. (1989). Perceived usefulness, perceived ease of use, and user acceptance of information technology. MIS Quarterly, 13(3), 319-340. https://doi.org/10.2307/249008

Denzin, N. K. (1978). The research act: A theoretical introduction to sociological methods (2nd ed.). McGraw-Hill.

Eastman, C., Teicholz, P., Sacks, R., & Liston, K. (2011). BIM handbook: A guide to building information modeling for owners, managers, designers, engineers, and contractors (2nd ed.). Wiley.

Eslamdoust, S., Lee, J. H., & Bohrani, T. (2024). Enhancing team performance in the digital age: Impact of technologically moderated communication in the interplay of e-leadership & trust. International Journal of Business & Management Studies, 5(04), 56-67.

Esmaeili, M., Ahmadi, M., Ismaeil, M. D., Mirzaei, S., & Canales Verdial, J. (2024). Advancements in AI-driven customer service. In 2024 IEEE World AI IoT Congress (AIIoT) (pp. 1-5). IEEE.

Feng, Z., Gao, Y., & Zhang, T. (2021). Gamification for visualization applications in the construction industry. In Industry 4.0 for the built environment: Methodologies, technologies and skills. Springer.

Ferizaj, D., Perotti, L., Dahms, R., & Heimann-Steinert, A. (2023). Use of technology in old age: Associations between acceptance, competence, control, interest and social indicators in individuals over 60 years old. Zeitschrift fur Gerontologie und Geriatrie, 57(3), 227-234.

Fernandes, D., Nikkel, M., & Guven, G. (2024). BIM-AI-VR integration for real-time model update and visualization. In Canadian Society of Civil Engineering Annual Conference (pp. 169-177).

Fowler, F. J. (2014). Survey research methods (5th ed.). SAGE.

Grassini, S., Thorp, S., Saevild Ree, A., Sevic, A., & Cipriani, E. (2025). Attitudes toward technology and artificial intelligence: The role of demographic and personality factors. Proceedings of the 36th Annual Conference of the European Association of Cognitive Ergonomics (EACE), 1-5.

Hargittai, E. (2002). Second-level digital divide: Differences in people’s online skills. *First Monday, 7*(4). https://doi.org/10.5210/fm.v7i4.942

Hosmer, D. W., Lemeshow, S., & Sturdivant, R. X. (2013). Applied logistic regression (3rd ed.). Wiley.

Hua, Y., Zhang, Y., & Ma, P. (2024). The mediating effect of BIM application on the link between organizational support and BIM user satisfaction. Architectural Engineering and Design Management, 20(3), 636-655. https://doi.org/10.1080/17452007.2024.2322507

Hyarat, E., Hyarat, T., & Sweis, G. (2022). Barriers to the implementation of building information modeling among AEC companies. Buildings, 12(2), 150. https://doi.org/10.3390/buildings12020150

Inguva, G. (2014). Differences for employees who use BIM/VDC in the construction workplace [Master's thesis, Colorado State University]. Mountain Scholar.

Jiang, H.-J., Cui, Z.-P., Yin, H., & Yang, Z.-B. (2021). BIM performance, project complexity, and user satisfaction: A QCA study of 39 cases. Advances in Civil Engineering, 2021, Article 6654851. https://doi.org/10.1155/2021/6654851

Jick, T. D. (1979). Mixing qualitative and quantitative methods: Triangulation in action. Administrative Science Quarterly, 24(4), 602-611. https://doi.org/10.2307/2392366

Jo, H., & Ahn, H. Y. (2024). Understanding digital engagement: Factors influencing awareness and satisfaction of digital transformation. Discover Computing, 27(1), 23.

Kazmi, S. A. B., & Irshad, H. (2025). The role of digital communication tools and work-life balance in enhancing employee productivity in hybrid work environments. Contemporary Journal of Social Science Review, 3(3), 1574-1585.

Khoshkonesh, A., Mohammadagha, M., & Ebrahimi, N. (2026). An uncertainty-aware 4D/5D digital-twin framework for cost estimation and probabilistic schedule control: Field and benchmark validation from a Texas mid-rise project. Discover Civil Engineering, 3, 127. https://doi.org/10.1007/s44290-026-00523-w

Kiburu, L., Boso, N., & Njiraini, N. (2023). Exploring how demographic factors influence consumer attitudes and technology usage. Serbian Journal of Management, 18(2), 353-365.

Kumari, J. (2024). Research on user participation in digital assisted technology. Journal of Research in Science and Engineering, 6, 48-53.

Leon, A. C., Davis, L. L., & Kraemer, H. C. (2011). The role and interpretation of pilot studies in clinical research. Journal of Psychiatric Research, 45(5), 626-629. https://doi.org/10.1016/j.jpsychires.2010.10.008

Mirzaei, S., Bogus, S. M., & Bunt, S. (2026). A human-centered approach to reframing job satisfaction in the BIM-enabled construction industry. Journal of Information Technology in Construction (ITcon), 31, 439–460. https://doi.org/10.36680/j.itcon.2026.020

Morris, M. G., & Venkatesh, V. (2000). Age differences in technology adoption decisions: Implications for a changing workforce. *Personnel Psychology, 53*(2), 375–403. https://doi.org/10.1111/j.1744-6570.2000.tb00206.x

Mukherjee, D., & Gopal, N. (2024). Impact of digital transformation on employee job satisfaction: A bibliometric analysis. International Journal of Bibliometrics in Business and Management, 3(2), 95-146.

Nanta, T. M., Noermijati, N., Rohman, F., & Hussein, A. S. (2025). The effect of digital touchpoint usage experience on customer loyalty mediated by digital engagement and customer satisfaction. Businesses, 5(1), 3.

Nezhad, M. H., Castro, F. E. V., Woolf, B., & Arroyo, I. (2024). Math teachers' in-class information needs and usage for effective design of classroom orchestration tools. In *European Conference on Technology Enhanced Learning* (pp. 299–314). Springer. https://doi.org/10.1007/978-3-031-72315-5_21

Noghabaei, M., Heydarian, A., Balali, V., & Han, K. (2020). A survey study to understand industry vision for virtual and augmented reality applications in design and construction. Automation in Construction, 119, 103356. https://doi.org/10.1016/j.autcon.2020.103356

Omidmand, P., & Ataei, S. (2026). Artificial intelligence in experimental approaches: Growth hacking, lean startup, design thinking, and agile. arXiv preprint arXiv:2603.20688. https://arxiv.org/abs/2603.20688

Omidmand, P., Dorri, R., Mozaffari, A., & Ataei, S. (2025). Artificial intelligence applications in lean startup methodology: A bibliometric analysis of research trends and future directions. arXiv preprint arXiv:2512.22164. https://arxiv.org/abs/2512.22164

Park, D. Y., Choi, J., Ryu, S., & Kim, M. J. (2022). A user-centered approach to the application of BIM in smart working environments. Sensors, 22(8), 2871. https://doi.org/10.3390/s22082871

Peng, H., Wang, X., Wu, H., & Huang, B. (2025). Human-computer interaction empowers construction safety management: Breaking through difficulties to achieving innovative leap. Buildings, 15(5), 771.

Spearman, C. (1904). The proof and measurement of association between two things. The American Journal of Psychology, 15(1), 72-101. https://doi.org/10.2307/1412159

Su, S., Zhong, R. Y., & Jiang, Y. (2025). Digital twin and its applications in the construction industry: A state-of-art systematic review. Digital Twin, 2, 15. https://doi.org/10.12688/digitaltwin.17664.3

Succar, B. (2009). Building information modelling framework: A research and delivery foundation for industry stakeholders. Automation in Construction, 18(3), 357-375. https://doi.org/10.1016/j.autcon.2008.10.003

Thabane, L., Ma, J., Chu, R., Cheng, J., Ismaila, A., Rios, L. P., Robson, R., Thabane, M., Giangregorio, L., & Goldsmith, C. H. (2010). A tutorial on pilot studies: The what, why and how. BMC Medical Research Methodology, 10, Article 1. https://doi.org/10.1186/1471-2288-10-1

Ugut, G. S. S., Slamet, H. W., Wuisan, D. S. S., Rahman, A. W. A., & Millah, S. (2025). Modeling user satisfaction in the Alodokter health application through IT-driven user engagement. In 2025 4th International Conference on Creative Communication and Innovative Technology (ICCIT) (pp. 1-7).

Van Tam, N., Quoc Toan, N., Dinh Quy, N., & Le Dinh Quy, N. (2021). Factors affecting adoption of building information modeling in construction project implementation. Cogent Business & Management, 8(1), 1918848. https://doi.org/10.1080/23311975.2021.1918848

Venkatesh, V., & Morris, M. G. (2000). Why don't men ever stop to ask for directions? Gender, social influence, and their role in technology acceptance and usage behavior. *MIS Quarterly, 24*(1), 115–139. https://doi.org/10.2307/3250981

Venkatesh, V., Morris, M. G., Davis, G. B., & Davis, F. D. (2003). User acceptance of information technology: Toward a unified view. MIS Quarterly, 27(3), 425-478. https://doi.org/10.2307/30036540

Wilson, D. A. (2018). Accepting the future: Comparing the adoption of technology by age cohorts [Doctoral dissertation, Wichita State University].

Yaverbaum, G. J. (1988). Critical factors in the user environment: An experimental study of users, organizations and tasks. MIS Quarterly, 12(1), 75–88.

Yigitbas, E., Nowosad, A., & Engels, G. (2023). Supporting construction and architectural visualization through BIM and AR/VR: A systematic literature review. In IFIP Conference on Human-Computer Interaction (pp. 145-166).

Zhang, T., Wang, Y., Zhou, X., Liu, D., Ji, J., & Feng, J. (2025). Intelligent human-computer interaction for building information models using gesture recognition. Inventions, 10(1), 5.